\documentstyle[twocolumn,epsf]{jpsj}
\def\runtitle{
One-Dimensional ${\bf S=1}$ Spin-Orbital Model
with Uniaxial Single-Ion Anisotropy}
\def\runauthor{Satoshi   {\sc Miyashita} and Norio {\sc Kawakami}
} 
\title{One-Dimensional ${\bf S=1}$ Spin-Orbital Model
with Uniaxial Single-Ion Anisotropy}
\author{Satoshi {\sc  Miyashita}
\footnote{E-mail : satoshi@tp.ap.eng.osaka-u.ac.jp}
 and Norio {\sc Kawakami}}
\inst{Department of Applied Physics,
Osaka University, Suita, Osaka 565-0871, Japan}

\recdate{\today}

\abst{%
We investigate ground-state properties of a
one-dimensional $S=1$ spin-orbital model 
with or without uniaxial single-ion anisotropy. 
By means of the density matrix renormalization group method, 
we compute the ground-state energy, the magnetization curves
and  the correlation functions. 
We discuss how the ground-state properties depend on the 
two exchange couplings for orbital and spin sectors. The phase diagram
obtained is compared with that for the $S=1/2$ model.
We also address the effect of uniaxial single-ion anisotropy.
}

\kword
{ $S=1$ spin-orbital model, one dimension, uniaxial single-ion anisotropy, 
}

\begin{document}
\sloppy
\maketitle

%%%%%%%%%%%%%%%%%%%%%%%%%%%%%%%%%%%%%%%%%%%%%%%%%%%%%%%%%%%%%%%%%%%
%%%%%%%%%%%%%%%%%%%%%%%%%%%%%%%%%%%%%%%%%%%%%%%%%%%%%%%%%%%%%%%%%%%
%%%%%%%%%                   　　　　  %%%%%%%%%%%%%%%%%%%%%%%%%%%%%
%%%%%%%%% 1.  Introduction  　　　　  %%%%%%%%%%%%%%%%%%%%%%%%%%%%%
%%%%%%%%%                   　　　　  %%%%%%%%%%%%%%%%%%%%%%%%%%%%%
%%%%%%%%%%%%%%%%%%%%%%%%%%%%%%%%%%%%%%%%%%%%%%%%%%%%%%%%%%%%%%%%%%%
%%%%%%%%%%%%%%%%%%%%%%%%%%%%%%%%%%%%%%%%%%%%%%%%%%%%%%%%%%%%%%%%%%%

\section{Introduction}

The interplay of spin and orbital degrees of freedom 
 has provided a variety of interesting 
phenomena in correlated electron systems. 
To understand the role played by spin and orbital fluctuations,
several different  spin-orbital models
have been studied extensively. As the simplest example
among others, the one-dimensional (1D) spin-orbital model
with SU(4) symmetry, which  describes a spin-1/2 system 
with two-fold orbital degeneracy, has been 
investigated.\cite{Affleck,Suther,Itakura,ueda} 
A slightly extended model, the 1D spin-orbital model 
with SU(2)$\otimes$SU(2) symmetry,\cite{Pati,Azaria,Yamashita,Itoi}
has also been studied
and the ground-state phase diagram has been established,
which consists of a variety of phases including gapful/gapless 
spin and orbital phases, etc.

Some transition metal oxides such as manganetes and vanadates, 
for which the Hund coupling plays an important role, 
have higher spins with degenerate orbitals. 
This naturally motivates us to extend the above spin-orbital model 
to  a model possessing higher spins.  A specific generalization
 of the model to the $S=1$ case has been proposed for vanadates, 
such as YVO$_3$, and investigated 
in detail.\cite{Khaliullin,Sirker,Miyashita,Shen,Horsch}
In particular, the competition of the 
orbital-valence-bond (OVB) solid phase  and 
spin-ferromagnetic phase has been clarified,
in accordance with some experimental findings
in the low-temperature quasi-1D phase of YVO$_3$.
The realization of the OVB phase has been suggested 
in neutron diffraction experiments.\cite{Ulrich}

Motivated by the  above hot topics, we study 
an SU(2)$\otimes$SU(2) extension of the $S=1$ spin-orbital model in 1D.
In contrast to the model proposed for YVO$_3$,\cite{Khaliullin}
 our aim is to capture generic features inherent in the $S=1$ spin-orbital
model, which is to be compared with the $S=1/2$ model.
\cite{Affleck,Itakura,ueda,Pati,Azaria,Yamashita,Itoi}
We also take into account the effects of  single-ion anisotropy, which 
may play an important role for $S=1$ systems.
We exploit the density matrix renormalization group (DMRG) 
method\cite{DMRG} and  investigate quantum phase transitions.
 By calculating
 the ground state energy and the spin/orbital
correlation functions for a given strength of single-ion anisotropy, 
we obtain the phase diagram in the plane 
of two exchange-coupling constants for spin and 
orbital sectors.

This paper is organized as follows. After a brief explanation of the
model in the next section,  we present the DMRG results
in $\S$ 3 for the ground-state energy, the
spin/orbital magnetization curves and  the
correlation functions, from which we determine
 the phase diagram. 
In $\S$ 4, we address the effects of uniaxial single-ion anisotropy.
A brief summary is given in $\S$ 5.

%%%%%%%%%%%%%%%%%%%
\section{Model}
%%%%%%%%%%%%%%%%%%%%

We consider an $S=1$ extension of the 1D spin-orbital model 
with uniaxial single-ion anisotropy $D$, which is characterized by
two coupling constants $A$ and $B$ related to
 the spin and orbital degrees of freedom. The Hamiltonian reads
%%%%%%%%%%%%%%%%%%%%%%%%%%%%%%%%%%%%%%%%%%%%%%%%%%%%%%%%%%%%%%%%%%%%%%%%%%%%%%%
\begin{eqnarray}
 {\cal H} &=& J \sum_{i} 
         \left[ 
           \left( {\bf S}_{i} \cdot {\bf S}_{i+1} + A \right)
           \left( {\bf T}_{i} \cdot {\bf T}_{i+1} + \frac{B}{4} \right)
         \right]
\nonumber \\
   && +    D   \sum_{i} \left(S_{i}^{z}\right)^2, 
\label{so-model}
\end{eqnarray}
%%%%%%%%%%%%%%%%%%%%%%%%%%%%%%%%%%%%%%%%%%%%%%%%%%%%%%%%%%%%%%%%%%%%%%%%%%%%%%%
where ${\bf S}_{\it i}$ is an $S=1$ spin operator at the $i$-th 
site and ${\bf T}_i$ is a $T=1/2$  pseudo-spin operator acting 
on the doubly-degenerate orbital degrees of freedom.
 $J$ controls the magnitude of the exchange couplings, 
which will be taken as the energy unit in the following discussions. 
 The
$D$-term represents uniaxial single-ion anisotropy,
which has been discussed in detail so far in the Haldane spin 
chain systems.

At a special point ($A=B=1$ and $D=0$), the symmetry is enhanced to
SU(2)$\otimes$SU(2). It is known that 
the ground state in this case is the orbital liquid with a small spin gap, 
which is called the OVB solid state.
\cite{Shen,Khaliullin,Sirker,Horsch,Miyashita}
Actually, the special model studied by Khaliullin {\it et al.} for 
 cubic vanadates\cite{Khaliullin,Sirker,Horsch,Miyashita} includes
 the above special case, at which our model 
coincides with  theirs.

The model (\ref{so-model}) is regarded as a natural extension of the 1D
$S=1/2$ spin-orbital
 model with SU(2)$\otimes$SU(2) symmetry,
\cite{Affleck,Itakura,ueda,Pati,Azaria,Yamashita,Itoi}
for which  the $D$-term is absent, and  $A \rightarrow A/4$
for the spin ($S=1/2$) part.  The comparison of these two models
may allow us to clarify the role of spin fluctuations on the spin-orbital
model.

%%%%%%%%%%%%%%%%%%%%%%%%%%%%%%%%%%%%%%%%%%%%%%%%%%%%%%%%%%%%%%%%%%%
\section{Ground State Properties without Anisotropy}
%%%%%%%%%%%%%%%%%%%%%%%%%%%%%%%%%%%%%%%%%%%%%%%%%%%%%%%%%%%%%%%%%%%

In this section, we investigate the ground-state properties of 
the Hamiltonian (\ref{so-model}) with $D=0$,
 and determine the zero-temperature  phase diagram.
We first notice that the spin- (orbital-) ferromagnetic state should
 be the ground state 
for $B\rightarrow-\infty$ ($A\rightarrow-\infty$).
Therefore, when we calculate the ground-state energy of 
the spin-(orbital-)ferromagnetic state, 
 we can fix $\langle{\bf S}_i\cdot{\bf S}_{i+1}
\rangle=1$ $(\langle{\bf T}_i\cdot{\bf T}_{i+1}\rangle=1/4)$. 

%%%%%%%%%%%%%%%%%%%%%%%%%%%%%%%%%%%%%%%%%%%%%%%%%%%%%%%%%%%%%%%%%%
%%%%%%%%%%%%%%%%%%%%%%%  Fig. 1  %%%%%%%%%%%%%%%%%%%%%%%%%%%%%%%%%
%%%%%%%%%%%%%%&%%% Energy for A=D=0  %%%%%%%%%%%%%%%%%%%%%%%%%%%%%
%%%%%%%%%%%%%%%%%%%%%%%%%%%%%%%%%%%%%%%%%%%%%%%%%%%%%%%%%%%%%%%%%%
%%%%%%%%%%%%%%%%%%%%%%%%%%%%%%%%%%%%%%%%%%%%%%%%%%%%%%%%%%%%%%%%%%%%%%%%%%%%
\begin{figure}[thb]
\begin{center}
\leavevmode \epsfxsize=110mm 
\epsffile{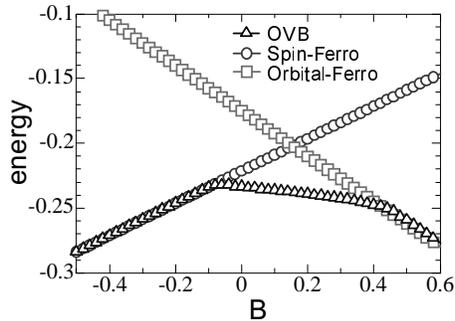}
\end{center}
\vskip -33mm
\caption{
The energy per site as a function of the exchange parameter $B$ for 
$A=D=0$; the ground-state energy obtained by the DMRG is shown 
by triangles.
We also plot the energy computed for  the spin-ferromagnetic state 
(circles) and the orbital-ferromagnetic 
state (squares). We clearly see 
two cusp structures, indicating  
 first-order phase transitions.  The intermediate region is 
identified as the OVB solid state.
}
%\vskip -20mm
\label{ene-a0}
\end{figure}
%%%%%%%%%%%%%%%%%%%%%%%%%%%%%%%%%%%%%%%%%%%%%%%%%%%%%%%%%%%%%%%%%%%

 In Fig. \ref{ene-a0}, the energy obtained by the DMRG 
is shown as a function of $B$ with keeping  $A=0$ fixed.
We find two first-order transition points; 
 $B_{c1}\simeq-0.09$ and  $B_{c2}\simeq0.41$. 
As clearly seen in Fig. \ref{ene-a0}, 
the system favors the spin-ferromagnetic state for small $B$ ($<B_{c1}$), 
but stabilizes the OVB solid state\cite{Khaliullin} 
for the intermediate region 
($B_{c1}<B<B_{c2}$). 
For large $B$ ($>B_{c2}$), the ground state is in 
the orbital-ferromagnetic phase. 

\begin{figure}[thb]
\begin{center}
\leavevmode \epsfxsize=110mm 
\epsffile{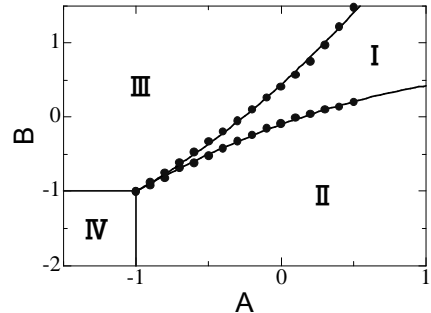}
\end{center}
\vskip -33mm
\caption{
Phase diagram of the 1D $S=1$ spin-orbital model
in the $A$-$B$ plane without uniaxial single-ion anisotropy.
The phase I is the OVB phase (see text). In the phase I$\!$I, the 
spin part is in the fully 
polarized state while the orbital part forms the gapless antiferromagnetic
 $T=1/2$ Heisenberg chain. On the other hand, in the phase I$\!$I$\!$I, 
the orbital sector is in the ferromagnetic phase while the spin sector is 
in the $S=1$ Haldane phase. Both of the spin and orbital parts are 
in the fully polarized ferromagnetic states in the phase I$\!$V. 
All the transitions are of first order.
}
\label{PD-D0}
\end{figure}

By repeating similar estimations of the critical points for other 
choices of $A$, we determine 
the ground-state phase diagram for $D=0$ in the $A$-$B$ plane,
which is shown in Fig. \ref{PD-D0}.
The phase diagram consists of
four phases; the OVB phase (I), 
the spin-ferromagnetic phase (I$\!$I), the orbital-ferromagnetic 
phase (I$\!$I$\!$I), and the spin-ferro/orbital-ferromagnetic phase (I$\!$V). 
We can estimate the phase boundary exactly in several limiting cases.
First, by dropping  the spin (orbital) degrees of freedom
when the spin- (orbital-) ferromagnetic ground state is stabilized, 
we immediately find that the phase boundaries of
I$\!$I-I$\!$V and I$\!$I$\!$I-I$\!$V 
 are exactly given by $A=-1$ and $B=-1$, respectively. 
We can also determine the exact asymptotic behavior of the boundary.
The boundary between the phases I and I$\!$I approaches 
$B=4|\epsilon_{g}^{(1/2)}|$ for $A\rightarrow\infty$, 
where $\epsilon_{g}^{(1/2)}= -\log 2 +1/4 \sim 
-0.443$ is the exact ground-state energy for 
the isotropic $T=1/2$ Heisenberg chain.\cite{bethe} 
On the other hand, for $B\rightarrow\infty$, the I-I$\!$I$\!$I 
boundary approaches $A=|\epsilon_{g}^{(1)}|$, where 
$\epsilon_{g}^{(1)}\sim-1.401484$ is the ground-state energy 
for the $S=1$ Haldane chain.\cite{DMRG} 
This asymptotic analysis is consistent with our phase diagram
obtained here numerically.

%%%%%%%%%%%%%%%%%%%%%%%%%%%%%%%%%%%%%%%%%%%%%%%%%%%%%%%%%%%%%%%%%%
%%%%%%%%%%%%%%%%%%%%%%%  Fig. 2  %%%%%%%%%%%%%%%%%%%%%%%%%%%%%%%%%
%%%%%%%%%%%%%%%%%  Phase Diagram for D=0  %%%%%%%%%%%%%%%%%%%%%%%%
%%%%%%%%%%%%%%%%%%%%%%%%%%%%%%%%%%%%%%%%%%%%%%%%%%%%%%%%%%%%%%%%%%
%%%%%%%%%%%%%%%%%%%%%%%%%%%%%%%%%%%%%%%%%%%%%%%%%%%%%%%%%%%%%%%%%%%%%%%%%%%%
%%%%%%%%%%%%%%%%%%%%%%%%%%%%%%%%%%%%%%%%%%%%%%%%%%%%%%%%%%%%%%%%%%%

%%%%%%%%%%%%%%%%%%%%%%%%%%%%%%%%%%%%%%%%%%%%%%%%%%%%%%%%%%%%%%%%%%
%%%%%%%%%%%%%%%%%%%%%%%  Fig. 3  %%%%%%%%%%%%%%%%%%%%%%%%%%%%%%%%%
%%%%%%%%%%%  Spin and Orbital Correlations for A=B  %%%%%%%%%%%%%%
%%%%%%%%%%%%%%%%%%%%%%%%%%%%%%%%%%%%%%%%%%%%%%%%%%%%%%%%%%%%%%%%%%
%%%%%%%%%%%%%%%%%%%%%%%%%%%%%%%%%%%%%%%%%%%%%%%%%%%%%%%%%%%%%%%%%%%%%%%%%%%%
\begin{figure}[thb]
\begin{center}
\leavevmode \epsfxsize=110mm 
\epsffile{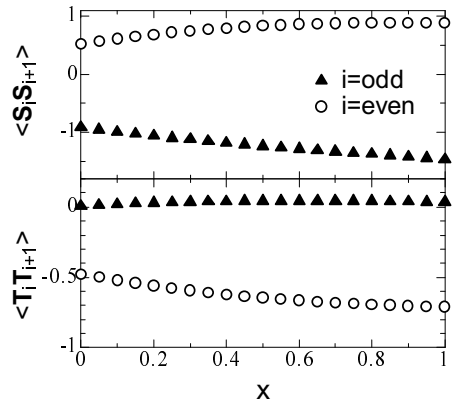}
\end{center}
\vskip -22mm
\caption{
Spin (upper panel) and orbital (lower panel) correlation functions 
in the phase I along the $A=B(= x)$ line in the OVB phase.
The site index $i$ is even (odd) for open circles (solid triangles). 
The even-odd $i$-dependence reflects a dimer property of the OVB phase. 
With increasing $x$, the dimerization gets strong and 
the size of a gap becomes large. 
}
\label{A=B-corre}
\end{figure}
%%%%%%%%%%%%%%%%%%%%%%%%%%%%%%%%%%%%%%%%%%%%%%%%%%%%%%%%%%%%%%%%%%%

%%%%%%%%%%%%%%%%%%%%%%%%%%%%%%%%%%%%%%%%%%%%%%%%%%%%%%%%%%%%%%%%%%
%%%%%%%%%%%%%%%%%%%%%%%  Fig. 4  %%%%%%%%%%%%%%%%%%%%%%%%%%%%%%%%%
%%%%%%%%%%%%%% Spin and Orbital Magnetization  %%%%%%%%%%%%%%%%%%%
%%%%%%%%%%%%%%%%%%%%%%%%%%%%%%%%%%%%%%%%%%%%%%%%%%%%%%%%%%%%%%%%%%
%%%%%%%%%%%%%%%%%%%%%%%%%%%%%%%%%%%%%%%%%%%%%%%%%%%%%%%%%%%%%%%%%%%%%%%%%%%%
\begin{figure}[thb]
\begin{center}
\leavevmode \epsfxsize=110mm 
\epsffile{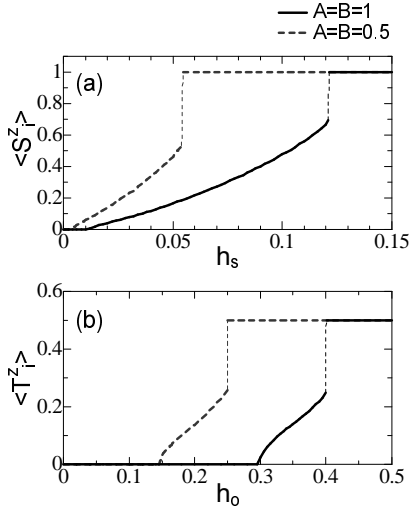}
\end{center}
\vskip -7mm
\caption{
 Spin (a) and  orbital (b) magnetization 
curves at $A=B=1$ (solid line) and 0.5 (dashed line). 
We calculate the spin (orbital) magnetization 
by setting  $h_o=0 (h_s=0)$. Note that both of the spin and orbital
magnetizations exhibit  typical behavior of  gapful systems, although
the orbital gap is much larger than the spin gap. For higher fields,
 the spin and orbital magnetizations show the first-order phase transition
 to the fully polarized state, which is signaled by a discontinuous jump. 
}
\label{smag&tmag}
\end{figure}
%%%%%%%%%%%%%%%%%%%%%%%%%%%%%%%%%%%%%%%%%%%%%%%%%%%%%%%%%%%%%%%%%%%

Although characteristic properties of the 
 phases I$\!$I, I$\!$I$\!$I and I$\!$V
can be directly deduced from those of the $S=1/2, 1$ chains,
 there appear nontrivial
properties due to the interplay of spins and
orbitals in  the OVB phase I.
We thus calculate the short-range spin and orbital 
correlation functions by taking  the line $A=B(= x)$
as representative parameters in the OVB phase.

The results are shown in Fig. \ref{A=B-corre}. 
As $x$ increases, the even-odd $i$-dependence of the 
correlation functions
 gets stronger both for spin and orbital sectors, implying
that the system favors the dimerization. 
At $x=1$, the orbital sector forms nearly perfect dimer singlets.
i.e. $\langle{\bf T}_i\cdot{\bf T}_{i+1}\rangle\sim -3/4$ for 
even $i$, while it is almost zero for odd $i$.
On the other hand,  two adjacent spins are almost parallel 
for even  $i$, i.e. $\langle{\bf S}_i\cdot{\bf S}_{i+1}\rangle\sim 1$,
 while for odd $i$, there are weak 
antiferromagnetic correlations.  These properties 
clearly characterize the dimer-like nature of the OVB 
phase.\cite{Khaliullin,Sirker,Miyashita}
It is remarkable that even if $x$ decreases from unity, the orbital
correlation $\langle{\bf T}_i\cdot{\bf T}_{i+1}\rangle$ for
odd $i$ stays very small, indicating that the orbital sector 
can be 
regarded as an assembly of approximately independent dimers 
irrespective of the values of $x$.

In order to observe the properties of the phase I in more detail, 
we compute the spin/orbital magnetization. 
To this end,
 we add the following terms to the Hamiltonian (\ref{so-model}), 
%%%%%%%%%%%%%%%%%%%%%%%%%%%%%%%%%%%%%%%%%%%%%%%%%%%%%%%%%%%%%%%%%
\begin{eqnarray}
{\cal H}_{ex} = - h_s \sum_i S_i^z - h_o \sum_i T_i^z, 
\end{eqnarray}
%%%%%%%%%%%%%%%%%%%%%%%%%%%%%%%%%%%%%%%%%%%%%%%%%%%%%%%%%%%%%%%%%
where $h_s$ and $h_o$ are external fields conjugate to
 the spin and orbital magnetizations. 
The calculated magnetization curves are shown in Fig. \ref{smag&tmag} 
for two typical choices of the  exchange couplings. 
Let us first observe  the spin magnetization 
for $x=A=B=1$ shown in Fig. \ref{smag&tmag}(a). 
The magnetization $\langle{S_i^z}\rangle$ is zero in small fields,
in accordance with the existence of a spin gap in the OVB phase. 
After the spin gap disappears at very low fields,
the magnetization increases smoothly, and jumps 
at the second critical point (e.g. $h_s\sim0.122$ for $x=1$),
driving the system to the spin-ferromagnetic state 
via  a first-order transition. 
Similarly, the orbital magnetization 
 shown in Fig. \ref{smag&tmag}(b) indicates that
the OVB state with $\langle{T_i^z}\rangle=0$
 is favored in small fields, and then the orbital magnetization
$\langle{T_i^z}\rangle$ develops gradually beyond
 a critical field  corresponding to the orbital gap.  
The system further undergoes
a first-order transition to the orbital-ferromagnetic state. 
Comparing the spin and orbital gaps, we notice that the orbital
gap is much larger than the spin gap,
\cite{Sirker,Miyashita}
consistent with the 
results for correlation functions, where the orbital-dimerization is 
quite strong.  In contrast to the low-field  behavior,
 both spin and orbital degrees of freedom change their
characters at the first-order
transition points. Note that the resulting high-field phase with
the fully spin (orbital) polarized state is the same as the 
 phase I$\!$I (I$\!$I$\!$I). 

%%%%%%%%%%%%%%%%%%%%%%%%%%%%%%%%%%%%%%%%%%%%%%%%%%%%%%%%%%%%%%%%%%
%%%%%%%%%%%%%%%%%%%%%%%  Fig. 5  %%%%%%%%%%%%%%%%%%%%%%%%%%%%%%%%%
%%%%%%%%%%%%%%%%% S=1/2 Phase diagram %%%%%%%%%%%%%%%%%%%%%%%%%%%%
%%%%%%%%%%%%%%%%%%%%%%%%%%%%%%%%%%%%%%%%%%%%%%%%%%%%%%%%%%%%%%%%%%
%%%%%%%%%%%%%%%%%%%%%%%%%%%%%%%%%%%%%%%%%%%%%%%%%%%%%%%%%%%%%%%%%%%%%%%%%%%%
\begin{figure}[thb]
\begin{center}
\leavevmode \epsfxsize=110mm 
\epsffile{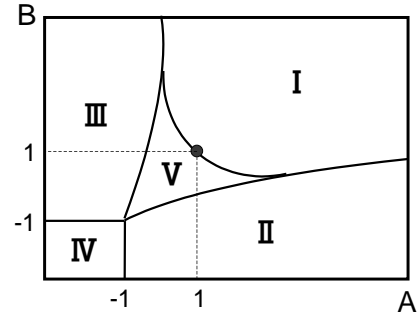}
\end{center}
\vskip -25mm
\caption{Phase diagram of the $S=1/2$ spin-orbital model.
\cite{Pati,Azaria,Yamashita,Itoi}
Phase I is a gapful phase with doubly degenerate ground states 
which form alternating spin and orbital singlets. 
In the phase I$\!$I, the spin degrees of freedom are in the fully
 polarized ferromagnetic state while the orbitals are in 
the antiferromagnetic ground state and vice versa 
in the phase I$\!$I$\!$I. 
Both spin and orbital degrees of freedom are in the fully polarized 
ferromagnetic states in the phase I$\!$V. 
The phase V is a gapless phase including an integrable SU(4) 
point ($A=B=1$).
}
\label{phase-half}
\end{figure}
%%%%%%%%%%%%%%%%%%%%%%%%%%%%%%%%%%%%%%%%%%%%%%%%%%%%%%%%%%%%%%%%%%%

Before closing this section, we compare the present 
phase diagram with that for the $S=1/2$ spin-orbital 
model,\cite{Pati,Azaria,Yamashita,Itoi}
which is sketched in Fig. \ref{phase-half}.
The corresponding spin-orbital Hamiltonian is given by 
(\ref{so-model}) by putting $S=1/2$ and replacing  
$A$ with $A/4$ for the spin sector.
Overall features are similar both in
  $S=1$ and $1/2$ models: the phases I$\!$I, I$\!$I$\!$I and
I$\!$V  are respectively
characterized by the spin-ferro/orbital-singlet state, the
spin-singlet/orbital-ferro state and the spin-ferro/orbital-ferro 
state.  The phase I also exhibits similar dimer-like
properties in both  cases. A remarkable difference is the gapless
phase V realized in the $S=1/2$ model around the
SU(4) point. This phase, which is stabilized by enhanced quantum fluctuations 
both in spin and orbital degrees of freedom,  
 disappears in the $S=1$ model.  We think that
the phase V is inherent in  the $S=1/2$ case, and
 a higher-spin extension of the SU(2)$\otimes$SU(2) model may
 have a phase diagram similar to the present $S=1$ case.

In contrast to the $S=1/2$ model, the single ion anisotropy
plays an important role for the $S=1$ model, which may provide a 
rich phase diagram. This problem is addressed in the following.

%%%%%%%%%%%%%%%%%%%%%%%%%%%%%%%%%%%%%%%%
\section{Effects of Uniaxial Single-Ion Anisotropy}
%%%%%%%%%%%%%%%%%%%%%%%%%%%%%%%%%%%%%%%%

\subsection{$D<0$}

Let us now consider the effects of uniaxial single-ion anisotropy. 
We start with  the ground-state properties for the case of $D<0$. 

Recall that in the phases I$\!$I and I$\!$V in Fig. \ref{PD-D0}, 
the spin sector is 
in the fully-polarized state, so that the nature of
these phases may not be altered upon the introduction of $D<0$,
except that  the direction of spin ordering is now fixed
to the $z$-axis. On the other hand, 
 the negative $D$  suppresses 
quantum fluctuations in the spin-gap phases I and I$\!$I$\!$I, 
resulting in 
 the antiferromagnetic spin order with $Z_2$ 
symmetry breaking. We refer to the corresponding  
new ordered phases
as the phase I$_m$  and I$\!$I$\!$I$_m$, respectively.

\begin{figure}[thb]
\begin{center}
\leavevmode \epsfxsize=110mm 
\epsffile{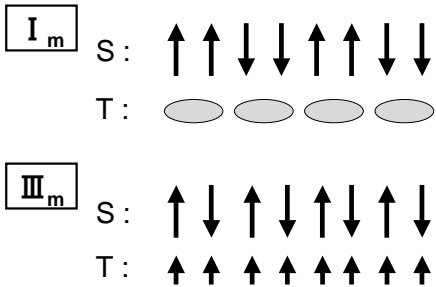}
\end{center}
\vskip -33mm
\caption{Schematic description  of spin orders for 
the phases I$_m$ and I$\!$I$\!$I$_m$. 
The label S (T)  denotes the spin (orbital) part, respectively.
}
\label{punch}
\end{figure}

Note that different types of spin order should
emerge in the phases I$_m$ and I$\!$I$\!$I$_m$.
The spin-antiferromagnetic order 
has the 4-site period (up-up-down-down alignment) in the phase I$_m$, and
the 2-site period (up-down-up-down alignment) in the phase I$\!$I$\!$I$_m$,
as schematically shown in Fig. \ref{punch}. 
This difference comes from the ground-state properties without 
anisotropy. Namely, in the OVB phase I, 
spins favor a ferromagnetic  configuration in a
 orbital-dimer singlet and
an antiferromagnetic  configuration between adjacent 
orbital singlets, so that
the spin sector is described by 
an effective ferro-antiferromagnetic bond-alternating chain. 
On the other hand, in the phase I$\!$I$\!$I$_m$, 
the spin sector forms the $S=1$ Haldane state at $D=0$, 
so that it naturally leads to an ordinary
 antiferromagnetic order upon the introduction of $D<0$.

%%%%%%%%%%%%%%%%%%%%%%%%%%%%%%%%%%%%%%%%%%%%%%%%%%%%%%%%%%%%%%%%%%
%%%%%%%%%%%%%%%%%%%%%%%  Fig. 6  %%%%%%%%%%%%%%%%%%%%%%%%%%%%%%%%%
%%%%%%%%%%%%%%%%% Schematic description %%%%%%%%%%%%%%%%%%%%%%%%%%
%%%%%%%%%%%%%%%%%%%%%%%%%%%%%%%%%%%%%%%%%%%%%%%%%%%%%%%%%%%%%%%%%%
%%%%%%%%%%%%%%%%%%%%%%%%%%%%%%%%%%%%%%%%%%%%%%%%%%%%%%%%%%%%%%%%%%%%%%%%%%%%
%%%%%%%%%%%%%%%%%%%%%%%%%%%%%%%%%%%%%%%%%%%%%%%%%%%%%%%%%%%%%%%%%%%

\begin{figure}[thb]
\begin{center}
\leavevmode \epsfxsize=120mm 
\epsffile{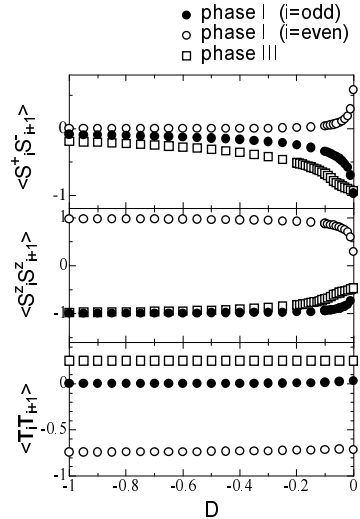}
\end{center}
\vskip -10mm
\caption{
Spin (upper and middle panels) and orbital (lower panel) correlation 
functions as a function of $D$:
$A=B=1$ (solid circles : $i=odd$ and open circles : $i=even$) and 
$A=-1$, $B=1$ (open squares). 
}
%\vskip -5mm
\label{corre-negaD}
\end{figure}

The above tendency to spin ordering  is 
indeed observed in the nearest-neighbor
correlation functions shown in
Fig. \ref{corre-negaD} as a function of $D(<0)$.
We exploit two typical values of
$A$ and $B$, which
correspond to the parameters for the phases I and I$\!$I$\!$I at $D=0$. 
Even in the presence of $D$, both 
spin and orbital correlation functions show (do not show)
the even-odd $i$-dependence for the parameters corresponding to 
the phase I$_m$ (I$\!$I$\!$I$_m$). It should be also noticed that 
the orbital correlations are almost independent of 
$D$, as should be naively expected:  
almost isolated orbital-dimer states are favored
in the phase I$_m$, while the orbital sector is always in the
fully polarized state,
 $\langle{\bf T}_i\cdot{\bf T}_{i+1}\rangle = 1/4$,
in the phase I$\!$I$\!$I$_m$ .

%%%%%%%%%%%%%%%%%%%%%%%%%%%%%%%%%%%%%%%%%%%%%%%%%%%%%%%%%%%%%%%%%%
%%%%%%%%%%%%%%%%%%%%%%%  Fig. 7  %%%%%%%%%%%%%%%%%%%%%%%%%%%%%%%%%
%%%%%%% Spin and Orbital Correlation for phase I&III %%%%%%%%%%%%%
%%%%%%%%%%%%%%%%%%%%%%%%%%%%%%%%%%%%%%%%%%%%%%%%%%%%%%%%%%%%%%%%%%
%%%%%%%%%%%%%%%%%%%%%%%%%%%%%%%%%%%%%%%%%%%%%%%%%%%%%%%%%%%%%%%%%%%%%%%%%%%%
%%%%%%%%%%%%%%%%%%%%%%%%%%%%%%%%%%%%%%%%%%%%%%%%%%%%%%%%%%%%%%%%%%%

As mentioned above, the spin-antiferromagnetic
order  with the 4 (2)-site period should be
stabilized  beyond a certain critical value of $|D|$
in the phase I$_m$ (I$\!$I$\!$I$_m$). Therefore,
 spin fluctuations are suppressed and thus
the correlation  
$\langle S^z_i \cdot S^z_{i+1} \rangle
\rightarrow \pm1$ while 
$\langle S^+_i \cdot S^-_{i+1}\rangle
\rightarrow 0 $ as $|D|$ increases. 
We now determine the phase transition point for
 I$\!$I$\!$I$\rightarrow$I$\!$I$\!$I$_m$. 
Note that the orbital sector always orders 
ferromagnetically,  so that 
the Hamiltonian (\ref{so-model}) for the spin sector is reduced to 
%%%%%%%%%%%%%%%%%%%%%%%%%%%%%%%%%%%%%%%%%%%%%%%%%%%%%%%%%%%%%%%%%%%%%%
\begin{eqnarray}
 {\cal H}_{\it OF} &=& J' \sum_{i} 
           \left( {\bf S}_{i} \cdot {\bf S}_{i+1} + A \right)
    +    D   \sum_{i} \left(S_{i}^{z}\right)^2,
\label{TF-model}
\end{eqnarray}
%%%%%%%%%%%%%%%%%%%%%%%%%%%%%%%%%%%%%%%%%%%%%%%%%%%%%%%%%%%%%%%%%%%%%%%%%%%%%%%
with $J'=J(1+B)/4$.  In this case,
the competition of the single-ion anisotropy $D$ and the 
effective exchange-coupling $J'$ results in the following transition,
which is independent of $A$. 
According to the results for the isotropic $S=1$ chain,
\cite{Tonegawa,WChen,Koga,Hikihara} 
the ground state is 
in the  Haldane phase for $-0.35J'{\leq}D{\leq}J'$, 
and in the Ne$\grave{e}$l phase for $D<-0.35J'$. 
Therefore, the critical value  is given by
 $D_c=-0.175J$ in Fig. \ref{corre-negaD},
which separates the Haldane phase  
and  Ne$\grave{e}$l phase for the spin sector.

It is more difficult  to determine the boundary between
I and I$_m$ only from our data for the correlation functions in 
Fig. \ref{corre-negaD}, partially because the spin gap 
is very small in the phase I.
Nevertheless, by taking into account that the spin gap is much
smaller than the orbital gap (Fig. \ref{smag&tmag}), we reasonably
expect that the corresponding critical value of $D$, which separates
I and I$_m$ in Fig. \ref{corre-negaD}, may be quite small,
say, less than $- 0.1$. The precise critical value should be 
 obtained by exploiting improved numerical methods, which 
we wish to leave for the future study.
 
\begin{figure}[thb]
\begin{center}
\leavevmode \epsfxsize=110mm 
\epsffile{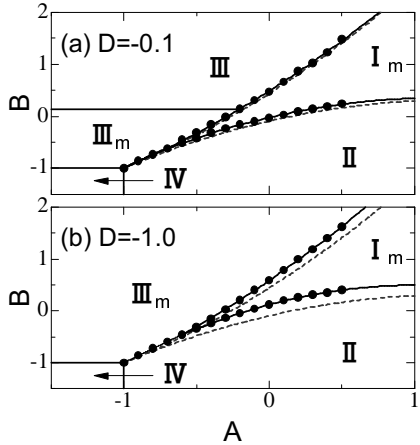}
\end{center}
\vskip -17mm
\caption{
Phase diagrams for $D<0$: (a)$D=-0.1$ and (b)$D=-1.0$.  The
phases labeled as I$\!$I, I$\!$I$\!$I and I$\!$V are the same as in
Fig. \ref{PD-D0}.  I$_m$ and I$\!$I$\!$I$_m$ are
spin-ordered phases (see text). 
The solid lines except for the I$\!$I$\!$I-I$\!$I$\!$I$_m$
 boundary  represent 
first-order transitions.  For reference, the phase boundary
at $D=0$ is shown by the dashed lines.
}
%\vskip -5mm
\label{negative-D}
\end{figure}

We show the phase diagram in the $A$-$B$ plane for $D < 0$ 
in Fig. \ref{negative-D}. Since the transition from the OVB phase 
to the other phases is
of first order, we determine the corresponding boundary by
comparing three-types of the energy among the OVB solid state, the 
spin-ferromagnetic state and the orbital-ferromagnetic state.
As mentioned above, the phase boundary between 
the phases I$\!$I$\!$I and I$\!$I$\!$I$_m$ is given by
%%%%%%%%%%%%%%%%%%%
\begin{eqnarray}
B_c \simeq  -11.4 D -1, 
\end{eqnarray}
%%%%%%%%%%%%%%%%%%%%%%%%%%%%%
which is independent of $A$.
For example,  $B_c\simeq 0.14$ for $D=-0.1$, as shown 
in Fig. \ref{negative-D}(a).
In the case of $D=-1.0$ in Fig. \ref{negative-D}(b),
 the phase I$\!$I$\!$I$_m$  dominates
the phase I$\!$I$\!$I in the displayed region,
because the boundary between these two phases is 
at $B_c\simeq 10.4$ for $D=-1.0$.
Note that the orbital sector orders ferromagnetically 
for all $D$ in  the phases I$\!$I$\!$I,
I$\!$I$\!$I$_m$ and  I$\!$V.
As mentioned above, the spin state in the OVB phase I is 
 sensitive to $D$,   so that 
 the ordered phase I$_m$ 
dominates the phase I both for $D=-0.1$ and $-1.0$
in the region shown in Fig. \ref{negative-D}

%%%%%%%%%%%%%%%%%%%%%%%%%%%%%%%%%%%%%%%%%%%%%%%%%%%%%%%%%%%%%%%%%%
%%%%%%%%%%%%%%%%%%%%%%%  Fig. 8  %%%%%%%%%%%%%%%%%%%%%%%%%%%%%%%%%
%%%%%%%%%%%%%% Phase Diagram for D<0  %%%%%%%%%%%%%%%%%%%%%%%%%%%%
%%%%%%%%%%%%%%%%%%%%%%%%%%%%%%%%%%%%%%%%%%%%%%%%%%%%%%%%%%%%%%%%%%
%%%%%%%%%%%%%%%%%%%%%%%%%%%%%%%%%%%%%%%%%%%%%%%%%%%%%%%%%%%%%%%%%%%%%%%%%%%%
%%%%%%%%%%%%%%%%%%%%%%%%%%%%%%%%%%%%%%%%%%%%%%%%%%%%%%%%%%%%%%%%%%%

%%%%%%%%%%%%%%%%%%%%%%%
\subsection{$D>0$}
%%%%%%%%%% %%%%%%%%%%%%%

Let us now move to the positive $D$ case. 
We start with  the phases  I$\!$I$\!$I and  I$\!$V,
for which   the phase transition points are 
easily determined, since the orbital sector
is in the fully polarized state.
In the phase I$\!$I$\!$I, the  so-called large-$D$ phase  is 
stabilized in the spin sector for $D>J'=J(1+B)/4$; i.e. for
$-1 <B<B_c$ ($B_c=-1+4D$), the large-$D$ phase appears,
while  the Haldane phase persists for $B>B_c$. On the 
other hand, in the phase I$\!$V, the XY phase emerges 
\cite{WChen,Schultz,Nijs}
in the region $-1-4D<B<-1$ in the presence of the $D$-term.

\begin{figure}[thb]
\begin{center}
\leavevmode \epsfxsize=110mm 
\epsffile{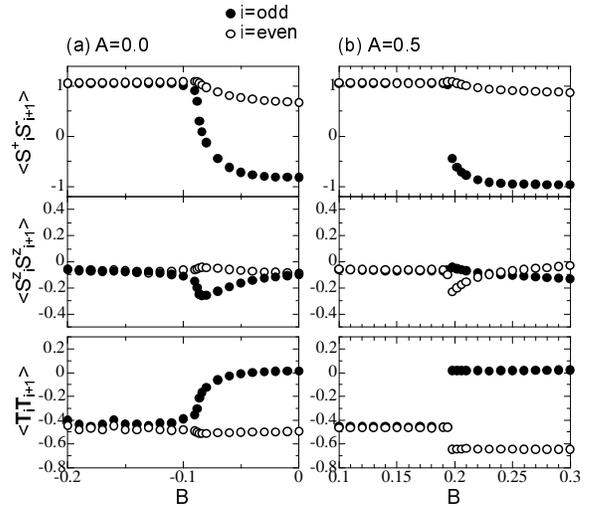}
\end{center}
\vskip -5mm
\caption{Spin (upper and middle panels) and orbital (lower panel) 
correlation 
functions for $D=0.01$ as a function of $B$ ((a) $A=0.0$, (b) $A=0.5$). 
The site index $i$ is even (open circle) or 
odd (solid circle). 
The even-odd $i$-dependence reflects dimer properties of the phase I. 
}
\label{corre-D001}
\end{figure}

It turns out that the boundary between the phases 
I and I$\!$I is quite
sensitive to positive $D$, and exhibits somewhat complicated features.
This is contrasted  to the robust first-order transition between 
the phases I and I$\!$I$\!$I, which hardly changes 
its character as far as  $D$ is small.
To make the above point clear, we
observe the behavior of the correlation functions 
by  choosing $D=0.01$ with two typical values of $A=0$ and 0.5.
In Fig. \ref{corre-D001} (a), 
the $B$-dependence of the spin  and orbital correlation functions 
are shown for $A=0$. 
For small $B$,  the ground state is in the phase I$\!$I, so that 
both spin and orbital correlation functions are spatially uniform
without  the even-odd $i$-dependence.
With increasing $B$, a continuous quantum phase transition occurs 
from the phase I$\!$I to I at $B_c\sim -0.1$, in contrast to the first-order 
transition at $D=0$, as seen in Fig. \ref{corre-D001}(a).
Beyond the critical value $B_c$, 
 the spin and orbital correlation functions exhibit
the even-odd $i$-dependence, reflecting dimer properties 
of the OVB solid state. 
Unfortunately, we cannot figure out  
whether this continuous transition is 
Berezinskii-Kosterlitz-Thouless (BKT) type \cite{Berezin,Kosterlitz,Kogut} 
or not only from the results shown in Fig. \ref{corre-D001},
because an extremely large system size is necessary to 
obtain the sensible results in our spin-orbital model
with a tiny spin gap. 
%%Nevertheless, we reasonably  expect that the transition 
%%is second-order, according to the fact that the Haldane spin systems
%%show BKT  transition in the presence of the positive $D$.????
Nevertheless, we can discuss some characteristic properties of each phase 
from the correlation functions. First, we note that
 the spin sector in the phase I$\!$I may be changed to the XY phase
upon introduction of $D$.\cite{WChen,Schultz,Nijs} 
This can be inferred from the spin correlation
functions shown in Fig. \ref{corre-D001}, which indeed show the 
preference of the XY phase: 
$\langle S^+_i S^-_{i+1}\rangle\sim 1$ and 
$\langle S^z_i S^z_{i+1}\rangle\sim 0$.
This observation is also supported by
the following consideration.
 Note first that the 
orbital correlation function is not affected by $D$
in the phase I$\!$I, and takes the constant 
value $\langle{\bf T}_i\cdot{\bf T}_{i+1}\rangle\sim\epsilon_g^{(1/2)}$ 
characteristic of the
$T=1/2$ orbital chain, so that we can focus on 
the remaining $S=1$ spin part.
According to the isotropic $S=1$ ferromagnetic chain,\cite{WChen,Schultz,Nijs} 
the first-order phase transition from the spin-ferromagnetic phase 
to the XY phase occurs upon introducing $D$,
%%and the large-$D$ phase is stabilized for $D{\geq}J'$ 
%%via the BKT transition.\cite{Berezin,Kosterlitz,Kogut} 
%%%%%%%%%%%%%%%%%
which is consistent with our numerical results for 
the correlation functions.
%%It is further seen from the spin and orbital correlation functions
%%in Fig. \ref{corre-D001}
%%that this XY phase undergoes a transition  to the phase 
%%with 4-site period, where the spin-directions are mainly in the XY
%%plane, but also have finite components for the $z$ axis.
%%%%%%%%%%%%%%%%%%%%%%%%%%%%%%%%%%%%%%%%%%%%%%%%%%%%%%%%%%%%%%%%%%
%%%%%%%%%%%%%%%%%%%%%%%  Fig. 9  %%%%%%%%%%%%%%%%%%%%%%%%%%%%%%%%%
%%%%%%%%%% Spin and Orbital Correlation for D=0.01 %%%%%%%%%%%%%%%
%%%%%%%%%%%%%%%%%%%%%%%%%%%%%%%%%%%%%%%%%%%%%%%%%%%%%%%%%%%%%%%%%%
%%%%%%%%%%%%%%%%%%%%%%%%%%%%%%%%%%%%%%%%%%%%%%%%%%%%%%%%%%%%%%%%%%%%%%%%%%%%
%%%%%%%%%%%%%%%%%%%%%%%%%%%%%%%%%%%%%%%%%%%%%%%%%%%%%%%%%%%%%%%%%%%

The above continuous transition is in  contrast to that for $A=0.5$
shown in Fig. \ref{corre-D001} (b), where
the system  exhibits the first-order phase transition
accompanied by a clear jump at $B_c\sim0.2$ both in the 
spin and orbital correlation functions.  This characteristic
behavior is essentially the same as that for $D=0$.

%When we consider only the spin part by discarding 
%orbital degrees of freedom, 
%from well-known $S=1$ chain results, 
%the transition from the XY-phase to the large-$D$ phase is a BKT-type and 
%the large-$D$ phase crossover to the dimer phase. 
%We naively think that the transition from the XY-phase to the dimer phase 
%should be BKT-type. On the contrary, 
%if only the orbital sector exists by dropping the spin degrees of freedom, 
%we should observe the second-order transition, 
%where a uniform $T=1/2$ Heisenberg chain undergoes the 
%transition to the $T=1/2$ dimer state.  

\begin{figure}[thb]
\begin{center}
\leavevmode \epsfxsize=110mm 
\epsffile{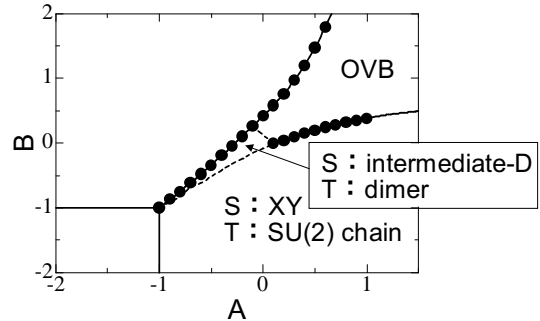}
\end{center}
\vskip -25mm
\caption{Phase diagram for the model with $D=0.01$.
The solid lines (with dots) indicate the first-order transition.
 Concerning the phase boundary
between I and I$\!$I,  there is a continuous transition 
(broken line)
for $A\leq0.1$, and a first-order transition (solid line with dots) 
for $A>0.1$.
The intermediate-$D$ phase is expected to emerge in the spin 
sector of the OVB phase for small $A$ and $B$.
Note also that around the 
boundary of $B=-1$, there appear two new spin phases:
the large-$D$ phase (XY spin phase) in the tiny region
$-1 <B<-0.96$ ($-1.04 <B<-1$), although not shown 
explicitly in the figure (see text). 
}
\label{phase-D001}
\end{figure}

By examining the correlation functions for other choices of $A$, 
we end up with a phase diagram expected for $D=0.01$, which is 
  shown in Fig. \ref{phase-D001}. 
The phase boundary between the phases I and  I$\!$I is
 either continuous or first-order  depending on the value of $A$:
 the transition is continuous
(first-order) for $A < 0.1$ ($A>0.1$). We have checked that
the critical value of $A$, which separates these
two types of  transitions, increases monotonically 
with the increase of $D$. For example, 
the critical value is  $A=0.5$ for  $D=0.012$.
For small $D$, the first-order transition 
dominates most part of the  I-I$\!$I phase boundary.

Here, an important question arises: what really
causes the change in the nature of the transitions,  first-order or 
 continuous, on the I-I$\!$I phase boundary.  To address this 
question, we should 
go into the detail of the phase I.
Since  the orbital part always forms the dimerized state 
in the phase I even for finite $D$, we again focus on the 
spin sector. Recall here  that
the spin sector in the OVB phase I forms the state 
similar to the $S=2$ Haldane-gap state (same universality
class).\cite{Shen,Sirker}  Fortunately,
the 1D $S=2$ Haldane system with uniaxial single-ion anisotropy 
was already investigated.\cite{Oshikawa,Scholl} 
Schollw\"ock numerically checked the disappearance of the spin gap 
at  a very small critical value of $D$ ($\sim 0.02 J$), and the system enters
the "intermediate-$D$ phase", which was claimed to be
 equivalent with the XY-phase.\cite{Scholl} 
Immediately after the above analysis, Oshikawa found that the behavior of 
the string correlation in $z$-axis deviates from a simple power-law, 
as a result, he suggested that this intermediate-$D$ phase is not 
truly gapless XY-phase, but somewhat modified one with 
some anomalous properties in the $z$-components. \cite{Oshikawa}
Anyway, the message from the above analysis is that the introduction of
small positive $D$ changes the Haldane phase to the 
so-called intermediate-$D$ phase.
In our  model, the size of the spin gap in the phase I gets small
as $A\rightarrow-1$ and $B\rightarrow-1$ (Fig. \ref{smag&tmag}).
Therefore, we think that the spin sector in the OVB phase is changed 
to the "intermediate-$D$ type phase" in the region with
 small $A$ and $B$, while the original spin-gap state in the OVB phase
 can still persist for 
larger $A$ and $B$. This may cause the change in the nature 
of the phase transition on the I-I$\!$I phase boundary.  
Unfortunately, our numerical 
calculation is not powerful enough to draw a definite 
conclusion about the existence of the  
 "intermediate-$D$ type phase".
We would like to clarify the detail in the future work.

%%%%%%%%%%%%%%%%%%%%%%%%%%%%%%%%%%%%%%%%%%%%%%%%%%%%%%%%%%%%%%%%%%
%%%%%%%%%%%%%%%%%%%%%%%  Fig. 10  %%%%%%%%%%%%%%%%%%%%%%%%%%%%%%%%
%%%%%%%%%%%%%% Phase diagram for D=0.01  %%%%%%%%%%%%%%%%%%%%%%%%%
%%%%%%%%%%%%%%%%%%%%%%%%%%%%%%%%%%%%%%%%%%%%%%%%%%%%%%%%%%%%%%%%%%
%%%%%%%%%%%%%%%%%%%%%%%%%%%%%%%%%%%%%%%%%%%%%%%%%%%%%%%%%%%%%%%%%%%%%%%%%%%%
%%%%%%%%%%%%%%%%%%%%%%%%%%%%%%%%%%%%%%%%%%%%%%%%%%%%%%%%%%%%%%%%%%

%%%%%%%%%%%%%%%%%%%%%%%%%%%%%%%%
\section{Summary}
%%%%%%%%%%%%%%%%%%%%%%%%%%%%%%

We have investigated quantum phase transitions of the 
1D $S=1$ spin-orbital model with uniaxial single-ion  anisotropy.
By means of the DMRG method, 
we have calculated the ground-state energy, the spin/orbital magnetizations,
 and the correlation functions, from which we 
have determined  the zero-temperature phase diagram. 

In the absence of the
 anisotropy, there appear four phases including the OVB phase
 characteristic of the $S=1$ spin-orbital system.
\cite{Khaliullin}
This phase has a small spin gap and  a fairly large orbital gap. 
In comparison with the $S=1/2$ spin-orbital model, we have found that
most of the phase diagram shows similar properties, but the 
phase V inherent in the $S=1/2$ model disappears in the $S=1$
case. We believe that a higher spin extension of the 
model should have the phase diagram similar to that of the
present $S=1$ model.

The introduction of uniaxial single-ion anisotropy $D$ 
gives rise to some interesting aspects. For negative $D$,
two different types of magnetic orders appear,
depending on whether the system at $D=0$ is in 
 the orbital-ferromagnetic phase or the OVB phase. 
Since the spin gap for the 
OVB phase is much smaller than that for
the orbital-ferromagnetic phase, the  
spin order in the OVB phase is induced even in the 
 small $|D|$ region.

On the other hand, it has turned out that the situation is
much more subtle in the positive $D$ case. 
In particular, the phase transition between the 
 OVB phase I and the spin-ferromagnetic phase I$\!$I
exhibits somewhat complicated feature:
the nature of transition changes from first-order to
continuous one,   depending sensitively on the value of $D$.
Although we have not been able to obtain the precise phase diagram, some
characteristic properties  of each phase have been discussed on the
basis of the correlation functions. These nontrivial
properties originate from  interplay of
the spin and orbital degrees of freedom, which may 
exemplify interesting aspects of the spin-orbital systems.
Further investigations should be done in the future study, especially
for the positive $D$ case.

%%%%%%%%%%%%%%%%%%%%%%%%%%%%%%%%%%%%%%%%
\section*{Acknowledgements}
%%%%%%%%%%%%%%%%%%%%%%%%%%%%%%%%%%%%%%%%
We would like to thank Giniyat Khalliulin and 
Akira Kawaguchi for fruitful discussions.
This work was partly supported by a Grant-in-Aid from the Ministry 
of Education, Science, Sports and Culture of Japan. 
A part of computations was done at the Supercomputer Center at the 
Institute for Solid State Physics, University of Tokyo
and Yukawa Institute Computer Facility. 

%%%%%%%%%%%%%%%%%%%%%%%%%%%%%%%%%%%%%%%%%%%%%%%%%%%%%%%%%%%%%%%%%%%%%
%                        REFERENCES                                 %
%%%%%%%%%%%%%%%%%%%%%%%%%%%%%%%%%%%%%%%%%%%%%%%%%%%%%%%%%%%%%%%%%%%%%
%

%%%

\end{document}